\documentclass[manuscript]{acmart}
\usepackage[normalem]{ulem}

\AtBeginDocument{%
  \providecommand\BibTeX{{%
    \normalfont B\kern-0.5em{\scshape i\kern-0.25em b}\kern-0.8em\TeX}}}

\begin{document}

%%
%% The "title" command has an optional parameter,
%% allowing the author to define a "short title" to be used in page headers.
\title[User Behavior Predicts Beliefs about Agents' Attributes]{My Actions Speak Louder Than Your Words: When User Behavior Predicts Their Beliefs about Agents' Attributes}

%%
%% The "author" command and its associated commands are used to define
%% the authors and their affiliations.
%% Of note is the shared affiliation of the first two authors, and the
%% "authornote" and "authornotemark" commands
%% used to denote shared contribution to the research.
% \author{Removed for Blind Review}
\author{Nikolos Gurney}
\email{gurney@ict.usc.edu}
\orcid{0000-0003-3479-2037}
\author{David V. Pynadath}
\email{pynadath@ict.usc.edu}
\orcid{0000-0003-2452-4733}
\author{Ning Wang}
\email{nwang@ict.usc.edu}
\affiliation{%
  \institution{Institute for Creative Technologies, University of Southern California}
  \streetaddress{12015 Waterfront Dr}
  \city{Los Angeles}
  \state{California}
  \country{USA}
  \postcode{90094}
}

%%
%% The abstract is a short summary of the work to be presented in the
%% article.
\begin{abstract}
An implicit expectation of asking users to rate agents, such as an AI decision-aid, is that they will use only relevant information---ask them about an agent's benevolence, and they should consider whether or not it was kind. Behavioral science, however, suggests that people sometimes use irrelevant information. We identify an instance of this phenomenon, where users who experience better outcomes in a human-agent interaction systematically rated the agent as having better abilities, being more benevolent, and exhibiting greater integrity in a post hoc assessment than users who experienced worse outcome---which were the result of their own behavior---with the same agent. Our analyses suggest the need for augmentation of models so that they account for such biased perceptions as well as mechanisms so that agents can detect and even actively work to correct this and similar biases of users.  
\end{abstract}

%%
%% The code below is generated by the tool at http://dl.acm.org/ccs.cfm.
%% Please copy and paste the code instead of the example below.
%%


%%
%% Keywords. The author(s) should pick words that accurately describe
%% the work being presented. Separate the keywords with commas.

%%
%% This command processes the author and affiliation and title
%% information and builds the first part of the formatted document.
\maketitle

\section{Introduction}
Perceived trustworthiness of an AI agent (we simplify this to \textit{agent}) is frequently referenced as an important determinant of successful human-agent interactions (e.g. \cite{wang2016trust, huang2017personal, ferrario2020ai}). A widely cited explanation for how humans think about trustworthiness posits that people consider three factors, or traits, of a person (or agent) when they evaluate trustworthiness: ability, benevolence, and integrity \cite{mayer1995integrative}. It is common practice for intelligent agent researchers to adapt a psychometric inventory of this three-factor model of trustworthiness for assessing users' perceived trustworthiness of agents \cite{mayer1999effect}. In theory, administering the inventory prior to an interaction allows researchers to assess the role of anticipated agent trustworthiness in users' behavior, while post hoc administration allows researchers to assess whether particular elements of an interaction, perhaps an experimental manipulation, impacted users' opinions of the agent. 

In practice, however, people frequently misuse information when they form judgments and make decisions \cite{kahneman1982judgment, gigerenzer2011heuristic}. For example, a person who is momentarily happy (sad), perhaps from reminiscing about a positive (negative) event from their recent past, is likely to rate their life satisfaction as higher (lower) than if you asked them when they were in a neutral state \cite{schwarz1983mood}. Regardless of the saliency of information, the normative approach is to always use it the same way. That is, the negative life event and the related emotions (the information) should always be factored into the assessment the same way---in most instances, because such information is highly subjective and often irrelevant, this means not integrating it at all. 

A person taking a normative approach to human-agent trust might form perceptions of the agent's ability, benevolence, and integrity based solely on its behavior, particularly the degree to which its behavior was consistent with that of someone who had those qualities. If people followed such a ``rational'' approach, we would expect their perceptions of the agent to depend entirely on its behavior and how said behavior impacted outcomes, not on their own subsequent decisions or feelings about those decisions regardless of the agent's input. To illustrate a \emph{non-normative approach}, imagine two graduate students writing literature reviews for a course on intelligent agents. The students are matched in every imaginable way, e.g. aptitude, intellect, motivation, etc., and both have access to the same intelligent, although fallible, literature-review agent. One student happens to appropriately follow the agent's correct advice and ignore its incorrect advice on relevant papers. The other student, who received the exact same advice from the agent, failed to make the right choices, purely by chance. Their respective success in navigating the literature is reflected in the grades that they receive, with the former student receiving a higher grade than the latter. A normative prediction would be that the grades received do not affect the students' perception of the agent, as the agent's advice to the students was identical. However, one can easily imagine that if both students rate the agent's traits after they are told their grades that the former student will rate the agent's traits more positively than the latter will, despite the agent providing them with the same input on their assignments and there being no meaningful difference between the students.   

More generally, we expect that people's perceptions of the agent's ability, benevolence, and integrity will be biased by the outcomes of the overall interaction, even when stochasticity and their own decisions affect those outcomes. In particular, the same agent exhibiting the same behavior, will be rated more positively on these measures by people who experience more positive outcomes, all else being equal. It is important to note that such a difference in ratings would appear even when people followed a Bayesian prescription \emph{if} the positive outcomes were attributable to better performance by the agent. To distinguish between the biased and unbiased perceptions, we must therefore ensure that the agent's behavior (including mistakes made, language used, etc.) is either identical across participants or else control for these differences in our analysis.

We report analyses testing this research question using data that were collected in two prior human-agent interaction studies, one of trust calibration \cite{wang2016trust} and another of compliance \cite{gurney2022trust,pynadath2022explainable}. Each data set represents interactions with a simulated robot that provided users with safety recommendations during a search-based task. After completing a mission, i.e., searching a fixed number of locations, study participants (users) responded to a battery of questions about their interactions with the agent, including a modified version of the trust inventory reported in \cite{mayer1999effect}. We show that a higher percentage of correct choices during the task was correlated with more positive ratings of the agent's traits, all else being equal. That is, after controlling for treatment condition and the total number of times that a participant complied with the robot's recommendations, participants percentage of correct compliance choices was still a meaningful instrument for predicting how they viewed the robot's ability, benevolence, and integrity ratings. 

We suggest that this points to the users' own private information, e.g. an affective state related to the experiment, coloring their evaluations of their AI teammate. Such behavior is well studied in humans and commonly known as the \textit{fundamental attribution error} \cite{ross1977intuitive}. This cognitive bias occurs when people overweight dispositional features of the person and underweight situational factors when forming judgments.
The results of our analyses are important for multiple reasons, but minimally because:
\begin{itemize}
    \item Agents can benefit from accurate models of the beliefs the people maintain about them \cite{gurney2022tom} and this result may prove beneficial to such models.
    \item It highlights a possible bias in the way that people think about agents which developers can potentially correct. 
    \item It suggests factors to be included in existing models of cognition that do not currently consider how such seemingly irrelevant personal information is used by people when they anthropomorphize an agent (e.g. \cite{reeves1996media,epley2007seeing}).
\end{itemize}

\section{Data}

\begin{figure*}
\caption{Screenshot from HRI testbed}
\label{fig:screenshot}
\includegraphics[width=.9\textwidth]{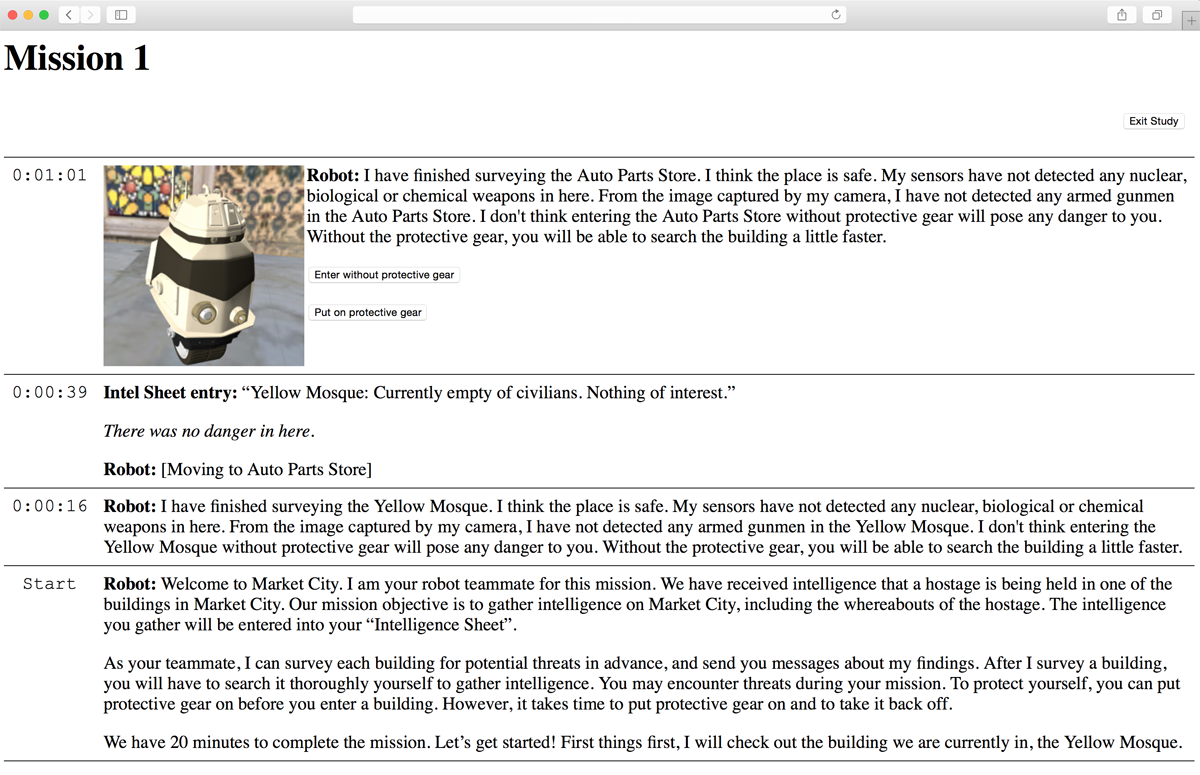}
\end{figure*}

Both experiments relied on adaptations of an online human-agent interaction simulation testbed from \cite{wang2015building}. The testbed was designed to study how giving an AI the ability to verbally explain its recommendations impacted participants' trust. This was accomplished by pairing participants with an intelligent robot during a simulated reconnaissance mission. Each mission was comprised of a series of potentially dangerous locations for the team to clear. Entering a dangerous location without the appropriate protective equipment always resulted in death, but the protective equipment also always neutralized a threat. The robot entered each location before the participant and then recommended whether or not the participant needed to put on protective gear. It based these recommendations on readings from three sensors---a camera, microphone, and nuclear, biological, and chemical (NBC) sensor---which were potentially faulty. The participants were then free to follow or ignore the robot's recommendation, with the outcome depending on their decision. During the second study, the robot also learned about the accuracy of its sensors using machine learning and then communicated to the study participant any updates that it made to its decision algorithm. Finally, the incentive structure was designed to motivate participants to clear locations quickly and without dying. Figure \ref{fig:screenshot} is a screenshot of the interface taken early in one such mission.

Importantly, the robot's sensor readings in both studies were scripted to generate a fixed sequence of correct/incorrect readings and recommendations. Thus, participants who experienced different outcomes did so only because of differences in their own decisions to follow or ignore the robot's recommendations.

\subsection{Study 1 Task and Treatment Conditions}

Participants in study 1 worked with one of eight different versions of the robot (a $2x4$ design) to complete three separate missions, each of which had eight locations to clear (see \cite{wang2016trust} for a complete explanation). Each participant worked with the same robot version for all three missions. All of the robots used a Partially Observable Markov Decision Process (POMDP) \cite{Kaelbling98} to form beliefs and select recommendations based on their sensor readings. The different robots, thus behaviors, were generated by crossing two ability levels, high and low, with four different types of explanations: confidence-level explanation, observation explanation based on either two or three sensors, and no explanation. A high-ability robot made the correct assessment of a location 100\% of the time. The low-ability robot had a faulty camera. This meant that the robot could not sense threats requiring visual detection (e.g. a hostile person), which resulted in it incorrectly reporting such locations as being safe for an overall accuracy of 80\%. With regard to the explanations, the robot always informed participants of its assessment of a location (e.g., ``I have finished surveying the opium den. I think the place is safe.''). Three treatment conditions augmented this with additional information at the end of the message: the robot's confidence in its assessment (e.g. ``I am 78\% confident about this assessment.'') or a statement of what it observed from either two or three sensors (e.g. ``My sensors have not detected any NBC weapons in here. From the image captured by my camera, I have not detected any armed gunmen in the cafe. My microphone picked up a friendly conversation.''). Note that when it reported sensor readings, it always reported the same type and number of readings for a participant.

Robot ability and explanation type had a significant main effect on a number of measured outcomes, including trust in the robot, beliefs about the transparency of its communications, compliance with its recommendations, and participants' fraction of correct decisions. Participants that worked with the high-ability robot were more positive in their ratings, but interestingly, they did not necessarily perform better given its ability. Also of note is the impact of explanations: in the case of the low ability robot, better explanations were correlated with better decision making by participants. More complete analyses and explanations are available in \cite{wang2016trust}. 

\subsection{Study 2 Task and Treatment Conditions}

Participants in study 2 worked with one of three robots during a single mission comprised of 45 locations \cite{gurney2022trust,pynadath2022explainable}. The robots of study 2 differed from those of study 1 in that they used model-free reinforcement learning (RL) \cite{Kaelbling96,sutton2018reinforcement} instead of POMDPs. The resulting dynamics in their behavior induced additional richness in their assessment explanations. The simplest robot offered no explanation of its assessment, which yielded messages similar to the base messages of study 1. The medium-sophistication robot added an explanation of its decision process (e.g. ``I think the place is dangerous. My NBC sensors have detected traces of dangerous chemicals.''). The most sophisticated explanation conveyed the robot's decision and learning processes (e.g. ``It seems that my estimate of the Unemployment Office was incorrect. I've updated the parameters of my model based on this result. I will give the same recommendation if my sensors pick up the same readings in the future.''). There was not an ability manipulation in study 2. 

The robot generated its decision and learning explanations based on self-generated decision-tree representations of its current RL policy. The robot used the path through the tree to explain its decision process and any changes to the tree during the most recent update of its policy to explain what it had learned \cite{pynadath2022explainable}. When the robot provided both a decision and learning explanation, participant compliance was significantly higher than when it provided only the decision or no explanation at all. Moreover, this effect appeared to be driven by the decision to follow the robot's advice to \textit{not} equip the protective gear. In the case of making correct decisions, the participants in the the decision and learning explanations group again did significantly better than the other two groups. More complete analyses and explanations are available in \cite{pynadath2022explainable}.

\subsection{Measures}
The main dependent variables of interest are the average responses to the three modified sub-measures for ability, benevolence, and integrity from the Mayer and Davis trust inventory. We are not interested here in predicting how mission outcomes impacted ``trust'' ratings as measured, holistically, by the Mayer and Davis scale items. Rather, we are interested in participants' perceptions of the individual traits that are measured by the scale and if these are predicted by their individual mission outcomes, all else being equal. 

The original scale was developed to measure trust in managers, which is reflected in the items (e.g. ``Top management is very capable of performing its job.''). Some items required only editing any references to management to refer to the robot instead. Others, however, were altogether irrelevant. The final measures used in the experiments are reproduced below with a ``*'' indicating reverse coding of a variable. 

In addition to the ability questions of the original Mayer and Davis inventory, the robot ability measure included items related to its specific instruments (camera, microphone, and NBC sensor).
\begin{enumerate}
    \item The robot is capable of performing its tasks.
    \item I feel confident about the robot's capability.
    \item The robot has specialized capabilities that can increase our performance.
    \item The robot is well qualified for this job.
    \item The robot's capable of making sound decisions based on its sensor readings.
    \item The robot's NBC (nuclear, biological and chemical weapon) sensor is capable of making accurate readings.
    \item The robot's microphone is capable of making accurate readings.
    \item The robot's camera is capable of making accurate readings.
    \item I feel confident about the robot's NBC sensor's sensing capability.
    \item I feel confident about the robot's camera's sensing capability.
    \item I feel confident about the robot's microphone's sensing capability.
    \item I feel confident about the robot's sensors.
    \item The robot is aware of its own limitations.
\end{enumerate}

The five benevolence items, again slightly modified from the original scale to better fit the human-agent interaction domain (versus management), are:
\begin{enumerate}
    \item The robot is concerned about my welfare. 
    \item I feel that my needs and desires are important to the robot.
    \item The robot would not knowingly do anything to hurt me.
    \item The robot looks out for what is important to me.
    \item The robot understands my goals in the mission.
\end{enumerate}
One item, ``Top management will go out of its way to help me'' was removed and replaced by item (5) because the robot's propensity to help participants was fixed, i.e., it could not go out of its way. 

Similarly, the integrity scale was trimmed from six to three items, but without replacement. Two removed items dealt with irrelevant topics (justice and fairness) and the third implied that the robot would share its plans, which it did not. The three items are:
\begin{enumerate}
    \item The robot's actions and behaviors are not very consistent.*
    \item I like the robot's values.
    \item Sound principles seem to guide the robot's behavior.
\end{enumerate}

Participants were asked to ``Please rate the extent to which you agree with the following statement'' for each item in the inventory. The 7-point scale ranged from \textit{Strongly Disagree} to \textit{Strongly Agree}. The measures for each sub-item are computed by mapping the scale responses to values from 1--7, 1 representing strongly disagree and 7 strongly agree, and averaging across the inventory items.

Participants in study 1 completed this inventory three times, once after each mission, thus we compute a composite score by averaging across all three measurements. Study 2 participants completed the inventory only once, so no composite was necessary.

Our independent variables include controls for the treatment conditions, which we delineated above and are addressed in the aforementioned citations. The two predictor variables of interest are the percentage of times that a participant followed the robot's recommendation and the percentage of times that they made the correct decision. In the three high-ability conditions of study 1, perfect compliance and perfect decision making coincide because the robot never errs. In the other three treatment conditions of study 1 and all the conditions of study 2, the robot made mistakes, thus perfect compliance did not equate to perfect decision making. 

\section{Analyses}
We rely on regression modeling to test our hypothesized relationship between participants' performance and their eventual responses to the inventory items. The control model is:
\begin{equation}
    Y_i = \beta_0 + \beta_\mathrm{Treat} X_\mathrm{i1} + \epsilon_i
\end{equation}
And the hypothesized model is:
\begin{equation}
    Y_i = \beta_0 + \beta_\mathrm{CP} X_\mathrm{iCP} + \beta_\mathrm{FP} X_\mathrm{iFP} + \beta_\mathrm{Treat} X_\mathrm{i1} + \epsilon_i
\end{equation}
$Y$ is the measure of a given trait, that is, ability, benevolence, or integrity. $\beta_0$ is the intercept and captures a control condition for the given experiment. These are the Constant values in the regression tables. $\beta_\mathrm{CP}$ is the variable of interest, correct percentage of choices. $\beta_\mathrm{FP}$ controls for the percentage of times that a participant followed the robot's advice. $\beta_\mathrm{Treat}$ captures the treatment conditions of the experiments, i.e., they account for what experimental condition that a participant experienced. We do not include the treatment condition fitted values in the tables for brevity's sake, as we are only interested in how a participant's choices during the mission(s) were correlated with their later ratings of the agent. The regression model, in effect, allows us to isolate the impact of the experimental conditions from the impact of a participant's own choices on their eventual ratings of the robot.

\begin{table*}[!htbp] \centering 
  \caption{Study 1 Regression Models} 
  \label{study1RegTab} 
\begin{tabular}{@{\extracolsep{1pt}}lcccccc} 
\\[-1.8ex]\hline 
\hline \\[-1.8ex] 
 & \multicolumn{6}{c}{\textit{Dependent variable:}} \\ 
\cline{2-7} 
\\[-1.8ex] & \multicolumn{2}{c}{Ability} & \multicolumn{2}{c}{Benevolence} & \multicolumn{2}{c}{Integrity} \\ 
\\[-1.8ex] & (1) & (2) & (3) & (4) & (5) & (6)\\ 
\hline \\[-1.8ex] 
 CP &  & 0.122 &  & 3.016$^{*}$ &  & 1.913$^{*}$ \\ 
  &  & (1.453) &  & (1.262) &  & (0.894) \\ 
  & & & & & & \\ 
 FP &  & 1.889 &  & $-$1.024 &  & $-$0.254 \\ 
  &  & (1.360) &  & (1.127) &  & (0.798) \\ 
  & & & & & & \\ 
 Constant & 5.871$^{***}$ & 4.344$^{***}$ & 4.557$^{***}$ & 2.741$^{***}$ & 4.466$^{***}$ & 3.036$^{***}$ \\ 
  & (0.144) & (0.677) & (0.227) & (0.621) & (0.161) & (0.440) \\ 
  & & & & & & \\ 
\hline \\[-1.8ex] 
Controls & Yes & Yes & Yes & Yes & Yes & Yes \\ 
Obs. & 78 & 78 & 199 & 199 & 199 & 199 \\ 
R$^{2}$ & 0.031 & 0.132 & 0.052 & 0.099 & 0.209 & 0.257 \\ 
Adjusted R$^{2}$ & 0.005 & 0.085 & 0.032 & 0.070 & 0.193 & 0.233 \\ 
\shortstack{Residual SE\\-} & \shortstack{0.777 \\(df = 75)} & \shortstack{0.745 \\(df = 73)} & \shortstack{1.355 \\(df = 194)} & \shortstack{1.328 \\(df = 192)} & \shortstack{0.965 \\(df = 194)} & \shortstack{0.941 \\(df = 192)} \\ 
\shortstack{$F$ Statistic \\ -} & \shortstack{1.203 \\(df = 2; 75)} & \shortstack{2.780$^{*}$ \\(df = 4; 73)} & \shortstack{2.654$^{*}$ \\(df = 4; 194)} & \shortstack{3.498$^{**}$ \\(df = 6; 192)} & \shortstack{12.837$^{***}$ \\(df = 4; 194)} & \shortstack{11.044$^{***}$ \\(df = 6; 192)} \\ 
\hline 
\hline \\[-1.8ex] 
\textit{Note:}  & \multicolumn{6}{r}{$^{*}$p$<$0.05; $^{**}$p$<$0.01; $^{***}$p$<$0.001} \\ 
\end{tabular} 
{\center Each column in the table represents a different regression model. (1), (3), and (5) are the control models in which the dependent variable of interest, noted above the column numbers, is predicted only by the treatment conditions of the experiment (the ``Controls,'' which are addressed in the original research). The other columns report the hypothesized models in which the correct choice percentage, CP, controlling for compliance, FP, predicts the trait ratings. The Constant $\beta$ parameter for each model is the average participant response to the given inventory in the basic experimental condition. The primary coefficient of interest, $\beta_\mathrm{CP}$, represents the change in the $\beta_\mathrm{Constant}$ when going from $0\%$ to $100\%$ correct decisions. The mean correct choices were 20.632 and 29.436 for studies 1 and 2, respectively. These are roughly what one would expect from the average participant captured by the Constant. Thus, adding the $\beta_\mathrm{CP}$ to this would be incorrect. Instead, adding $\beta_\mathrm{CP}/n$, where $n$ is the number of choices in the given study, yields the effect of the average person making one more correct choice (or one fewer when subtracted). \par}
\end{table*} 

\begin{table*}[!htbp] \centering 
  \caption{Study 2 Regression Models} 
  \label{study2RegTab} 
\begin{tabular}{@{\extracolsep{1pt}}lcccccc} 
\\[-1.8ex]\hline 
\hline \\[-1.8ex] 
 & \multicolumn{6}{c}{\textit{Dependent variable:}} \\ 
\cline{2-7} 
\\[-1.8ex] & \multicolumn{2}{c}{Ability} & \multicolumn{2}{c}{Benevolence} & \multicolumn{2}{c}{Integrity} \\ 
\\[-1.8ex] & (1) & (2) & (3) & (4) & (5) & (6)\\ 
\hline \\[-1.8ex] 
 CP &  & 3.034$^{*}$ &  & 4.522$^{**}$ &  & $-$0.707 \\ 
  &  & (1.377) &  & (1.640) &  & (1.605) \\ 
  & & & & & & \\ 
 FP &  & $-$2.884$^{*}$ &  & $-$5.211$^{***}$ &  & 1.130 \\ 
  &  & (1.271) &  & (1.513) &  & (1.481) \\ 
  & & & & & & \\ 
 Constant & 5.117$^{***}$ & 5.159$^{***}$ & 5.135$^{***}$ & 5.800$^{***}$ & 4.333$^{***}$ & 4.021$^{***}$ \\ 
  & (0.116) & (0.268) & (0.143) & (0.320) & (0.134) & (0.313) \\ 
  & & & & & & \\ 
\hline \\[-1.8ex] 
Controls & Yes & Yes & Yes & Yes & Yes & Yes \\ 
Obs. & 159 & 159 & 159 & 159 & 159 & 159 \\ 
R$^{2}$ & 0.150 & 0.177 & 0.025 & 0.114 & 0.012 & 0.023 \\ 
Adjusted R$^{2}$ & 0.139 & 0.156 & 0.013 & 0.091 & $-$0.0002 & $-$0.003 \\ 
\shortstack{Residual SE\\-} & \shortstack{0.861 \\(df = 156)} & \shortstack{0.852 \\(df = 154)} & \shortstack{1.058 \\(df = 156)} & \shortstack{1.015 \\(df = 154)} & \shortstack{0.992 \\(df = 156)} & \shortstack{0.993 \\(df = 154)} \\ 
\shortstack{$F$ Statistic \\ -} & \shortstack{13.737$^{***}$ \\ (df = 2; 156)} & \shortstack{8.297$^{***}$\\ (df = 4; 154)} & \shortstack{2.005\\ (df = 2; 156)} & \shortstack{4.966$^{***}$ \\(df = 4; 154)} & \shortstack{0.986 \\(df = 2; 156)} & \shortstack{0.899 \\(df = 4; 154)} \\ 
\hline 
\hline \\[-1.8ex] 
\textit{Note:}  & \multicolumn{6}{r}{$^{*}$p$<$0.05; $^{**}$p$<$0.01; $^{***}$p$<$0.001} \\ 
\end{tabular} 
\end{table*} 

We report the fitted values of the regression models in a separate table for each study. As noted, we do not report values for the treatment conditions, which are discussed in the original papers; however, their presence in the models is indicated by ``Controls'' and a ``Yes.'' Unfortunately, in Study 1, only a subset of participants provided complete responses to the Ability measure leaving the related models under powered (models (1) and (2) of Table \ref{study1RegTab}). The coefficients for two continuous variables, $\beta_\mathrm{CP}$ and $\beta_\mathrm{FP}$, should be interpreted as the expected percentage change in the dependent variable, all else being equal, if the predicted variable (a rating for Ability, Benevolence, or Integrity) were to go from zero to one.  

Interpretation of the $\beta_\mathrm{CP}$ values is not entirely straightforward. Recall that in the ordinary least squares regression model a $\beta$ value represents a one-unit change in the variable. Because our independent variables are percentages, a one unit change would represent going from no correct choices to all correct choices. The regression model, however, represent a hypothetical average participant and the mean correct choices were 20.632 and 29.436 for studies 1 and 2, respectively. Thus, to understand how getting one additional choice correct would hypothetically impact a benevolence rating, we can divide the $\beta_\mathrm{CP}$ value by the total number of choices and add or subtract this from the average benevolence rating to see how a 1 choice difference would, hypothetically, change the benevolence score. In study 1, the average increase in rated benevolence for each additional correct choice was $3.016/24 = 0.126$ and in study 2, $4.522/45 = 0.100$, all else being equal. With these explanations of the fitted values in mind, interpreting the models becomes much more meaningful. \footnote{CP and FP are, obviously, correlated, however, controlling for the variance that they mutually explain via an interaction term was not warranted. Doing so for both study 1 and study 2 did not result in better model fits, so we stick to the simpler models in which they only individually explain variance.}

In the case of both sets of study data, we reject the null hypothesis that there is no correlation between the correct choice percentage and ratings of an agent's benevolence (see the fourth column, first row of Tables \ref{study1RegTab} and \ref{study2RegTab}).  The more correct choices that participants made, the higher they rated the robot's benevolence. Keep in mind that these models are controlling for any differences in what the robot actually did or said. Moreover, \textit{F} tests from ANOVAs comparing the treatment only and hypothesized models support acceptance of the hypothesized models (see Table \ref{BenModComp}). This finding supports the idea that participants were using, arguably irrelevant, personal information when evaluating the robot's benevolence. When we control for the fraction of times that a participant followed the robot and the experimental treatment condition they received, the only experiential differences, then any remaining variance in the model explained by CP must be an artifact of the participants. To reiterate our position, if participants were only using the normatively correct information (the robot's recommendations, whether those recommendations were correct, and the harm created by bad recommendations), then the percentage of correct choices that a participant made should not factor into their evaluation of the robot. When it does, then it points to participants using extemporaneous information---we argue their own affective state, or how they feel about the interaction and outcome.

\begin{table}[!htbp] \centering 
  \caption{Benevolence Model Comparisons} 
  \label{BenModComp} 
\begin{tabular}{@{\extracolsep{2pt}}lcc} 
\\[-1.8ex]\hline 
\hline \\[-1.8ex] 
Statistic & \multicolumn{1}{c}{\shortstack{Study 1 \\ Control Model (3) Vs. CP Model (4)}} & \multicolumn{1}{c}{\shortstack{Study 2 \\ Control Model (3) Vs. CP Model (4)}}\\ 
\hline \\[-1.8ex] 
Sum of Sq & 17.532 & 15.963 \\ 
$F$ & 4.968 & 7.755 \\ 
Pr(\textgreater F) & 0.008 & <0.001 \\  
\hline \\[-1.8ex] 
\end{tabular} 
\end{table} 

% \begin{table}[!htbp] \centering 
%   \caption{Study 1 \\Benevolence Model Comparison} 
%   \label{stdy1BenModComp} 
% \begin{tabular}{@{\extracolsep{2pt}}lcc} 
% \\[-1.8ex]\hline 
% \hline \\[-1.8ex] 
% Statistic & \multicolumn{1}{c}{Control Model (3) Vs. CP Model (4)} \\ 
% \hline \\[-1.8ex] 
% Sum of Sq & 17.532 \\ 
% F & 4.968 \\ 
% Pr(\textgreater F) & 0.008 \\  
% \hline \\[-1.8ex] 
% \end{tabular} 
% \end{table} 

% \begin{table}[!htbp] \centering 
%   \caption{Study 2 \\Benevolence Model Comparison} 
%   \label{stdy2BenModComp} 
% \begin{tabular}{@{\extracolsep{2pt}}lcc} 
% \\[-1.8ex]\hline 
% \hline \\[-1.8ex] 
% Statistic & \multicolumn{1}{c}{Control Model (3) Vs. CP Model (4)} \\ 
% \hline \\[-1.8ex] 
% Sum of Sq & 15.963 \\ 
% F & 7.755 \\ 
% Pr(\textgreater F) & <0.001 \\ 
% \hline \\[-1.8ex] 
% \end{tabular} 
% \end{table} 

As noted, the Ability measure was only completed by a fraction of the participants in Study 1, leaving these models under powered. Study 1 included six treatment conditions. There are simply too few responses (78 in total that completed the measure) in each condition to have statistical power, nevertheless, we included the models for completeness. The data from Study 2 do suggest, however, that we can reject the null hypothesis that there is no correlation between the correct choice percentage and ratings of the agent's ability (columns (1) and (2) of Table \ref{study2RegTab}). Note, however, that the richer model, Table \ref{study2RegTab} column (2), did not explain a significantly higher amount of variance than the control model in column (1). Table \ref{stdy2AbiModComp} presents the F test from an ANOVA comparing these two models. We interpret this as weak support for our hypothesis that people are using personal information when evaluating the robot's ability. This may be attributable to the task setting and somewhat limited interactions with the robot. Whereas benevolence is a dispositional trait of the robot, ability is situational. It is feasible that participants were somewhat more reserved if they thought the robot would be better (or worse) if another setting, thus were less extreme in their evaluations.  

\begin{table}[!htbp] \centering 
  \caption{Study 2 \\Ability Model Comparison} 
  \label{stdy2AbiModComp} 
\begin{tabular}{@{\extracolsep{2pt}}lcc} 
\\[-1.8ex]\hline 
\hline \\[-1.8ex] 
Statistic & \multicolumn{1}{c}{Control Model (1) Vs. CP Model (2)} \\ 
\hline \\[-1.8ex] 
Sum of Sq & 3.747 \\ 
$F$ & 2.580 \\ 
Pr(\textgreater F) & 0.079 \\ 
\hline \\[-1.8ex] 
\end{tabular} 
\end{table} 

The data from Study 1 suggest that we can reject the null hypothesis that there is no correlation between the correct choice percentage and the ratings of the agent's integrity; however, the Study 2 data do not support this conclusion (columns (5) and (6) of Tables \ref{study1RegTab} and \ref{study2RegTab}). Additionally, an \textit{F} test from an ANOVA reported in Table \ref{stdy1IntModComp} revealed that the model which controls for correct choice percentage, Table \ref{study1RegTab} column (6), explains a significantly greater amount of variance than the control model, column (5). The integrity sub-scale, as we mentioned, was significantly trimmed due to many items in the scale lacking relevance for our setting. In retrospect, it is not surprising that these results were not significant as having fewer items in the measure will result in a more chaotic variance structure (i.e. make it less friendly to prediction).

\begin{table}[!htbp] \centering 
  \caption{Study 1 \\Integrity Model Comparison} 
  \label{stdy1IntModComp} 
\begin{tabular}{@{\extracolsep{2pt}}lcc} 
\\[-1.8ex]\hline 
\hline \\[-1.8ex] 
Statistic & \multicolumn{1}{c}{Control Model (1) Vs. CP Model (2)} \\ 
\hline \\[-1.8ex] 
Sum of Sq & 10.817 \\ 
$F$ & 6.108 \\ 
Pr(\textgreater F) & 0.003 \\ 
\hline \\[-1.8ex] 
\end{tabular} 
\end{table} 
% do we want to say something about FP??? The control parameter, $\beta\mathrm{FP}$, was also significant in study 2 and the effect was directional consistent in study 1, but failed to achieve significance ($p = .365$). 

\section{Discussion}
Psychologists have long studied how people ascribe individual factors, from knowledge (ability) to personality traits (benevolence and integrity), to others. Generally, those others are humans. Occasionally, they are animals, plants, and even inanimate objects. In these instances we say that the person is anthropomorphizing the object, or in other words, they are implying that the object has human-like traits. Increasingly, people interact with intelligent agents imbued with capabilities that reflect the factors we see in other humans, such as the robots of the above experiments which were capable of explaining their decision making processes in human-interpretable terms. Thus, it is somewhat unsurprising when people start ascribing those factors to such agents, just as they do with pets, etc. 

When a person anthropomorphizes an agent, they attribute human-like mental states to the agent which can lead them to react differently to its behavior \cite{epley2007seeing}. Researchers and developers that work in the HCI, HRI, and similar spaces explicitly try to curate such anthropomorphic responses \cite{zlotowski2015anthropomorphism,de2016almost}, whether through clever design (e.g. having an agent use gestures that match its speech \cite{salem2013err}) or actual implementation of human-like traits (e.g. incorporating personality traits into an agent that mimic those of a human \cite{cafaro2016first}). There is some evidence, albeit conflicted, that the personality traits designed into an agent may color how a person interprets said agent's behavior. For example, some people may or may not prefer that an agent's degree of extroversion matches their own \cite{isbister2000consistency,tapus2008user,cafaro2016first}. It is also worth noting that anthropomorphism is not necessarily a mindful process, meaning that the simple presence of cues that suggest human-like traits can result in mindless anthropomorphism of an agent \cite{kim2012anthropomorphism}. It is even possible that the simple act of asking people whether or not an agent acted in a benevolent fashion triggers them to anthropomorphise it, as it is well documented that they way in which we ask such questions can in part determine responses \cite{schwarz1999self}.

In psychology, the study of how a person gathers and uses information in causal judgments is known as \textit{attribution theory} \cite{fiske1991social}, a process that is closely tied to the tendency to anthropomorphize non-human objects. Psychologists generally divide attributions into dispositional, in which the cause of the behavior is assigned to the agent, and situational, in which the cause of behavior is assigned to events or other external factors. One of the earliest experimental investigations of attribution asked study participants to watch an animation in which shapes moved around on the screen and describe what they observed \cite{heider1944experimental}. The famous result of the experiment is that participants spontaneously attributed traits to the shapes solely on the short animation. The attributed traits included both dispositional traits, such as one shape being a bully, and situational, such as one shape being afraid of a particular event (but not having a generalized trait of fearfulness). The reconnaissance mission data demonstrate instances of both types of attribution: the agent's ability being situational and its degree of benevolence and integrity dispositional. Absent from the literature on anthropomorphizing agents and attribution theory, however, is the artifact observed in the reconnaissance mission data: individual's choices coloring the their attribution of traits to other agents. 

The observed effect appears to be a close cousin of a well-studied phenomenon in social psychology: the fundamental attribution error, or the tendency of people to overweight dispositional factors, like personality traits, and underweight situational factors when judging others behavior and the associated outcomes \cite{ross1977intuitive}. In interpersonal interactions, this often looks like blaming a poor outcome on a person, particularly because of some aspect of who they are, rather than contextual features that were more likely the causal factors. Unfortunately, the data that we analyzed do not support identification of the precise cognitive process that is driving the error that we observed. A few possible explanations include: using information about one's affective state (a person feels good or bad and project that onto the assessment), explanation-seeking behavior (a person needs to explain a phenomenon and it is more comfortable to blame the robot than themselves), or switching between causal and counterfactual reasoning based on outcomes, to name a few. In other words, situational factors, e.g. a person's affective state, can also influence personality trait judgments. In cases of repeated interactions with an agent, this sets up the potential for a perverse cycle where the accuracy of people's beliefs related to the factors engineered into an agent are contingent on, and reinforced by, their subjective experiences. There is an extensive literature on the fundamental attribution error, including work on how to reduce or eliminate the bias, in the literature. We believe that extending the existing literature into the human-agent systems is important future work.  
%Here, however, it appears that people are importing irrelevant information into their judgments of the agent's attributes. 
%---perhaps their own emotional state, which results from the mission outcome---

%\subsection{Implications}
%Aside from raising an intriguing set of questions about human judgment and decision making, we believe that this result is incredibly important for the IVA community...

\subsection{Limitations}
%Note: potential limitation of the integrity  scale. ---at end of section now
This research has three main limitations: 1) it considers ratings of only a narrow set of attributes, 2) it looks at data from only a single experimental paradigm, and 3) it does not consider the full spectrum of judgments made during trait attributions. Obviously, both humans and agents have a broad range of factors that define who they are. Some of these factors are situational, such as ability. If any of the authors were judged on their ability to play American football, for example, the reality is that we are poor players to such a degree that, even if a fellow teammate were judging us after winning the Super Bowl, they would almost certainly still say that we are terrible. Conversely, if one of us were asked to rate another on their research skills, the outcome of a recent journal or conference submission may very well color that assessment. On the other hand, certain ratings of some dispositional traits, such as integrity or benevolence, may be impacted by outcomes in domains where the factor under consideration does not matter. To illustrate, rather than rating research skills, perhaps one author rates the likeability of a coauthor after a paper rejection. Our results suggest that this rating would be lower, all else being equal, than if we just received an acceptance. 

This all, of course, is contingent on the result generalizing outside of the research paradigm that we examined. Although our intuition is that it will, we do not have data to support that intuition. Like any correlational result, it is entirely possible that the observed effect is an artifact of the particular experimental interactions and settings. Thus, additional research is needed to generalize the effect that we observe across different agent factors and settings. 

Finally, there are at least three different judgments a person must make when attributing agent factors as they did in the data we analyzed: is the attribute situationally relevant, whether the agent does in fact possess the attribute, and if so, to what degree. Each individual judgment comes with an opportunity to appropriately, or not, use information. Our interpretation of the result is that study participants misused their own, private information, e.g., their affective state related to mission performance, when making a judgment about the degree to which the agent possessed each trait. The first judgment, the situational relevance, was largely unnecessary as the questions were designed to either be or imply relevance. The second judgment, about the actual presence of the factors, was obviated as well by the questions: they were conveyed in a way that assumed that the agent, at least to some degree, had the attributes. Thus, the data that we analyzed did not allow us to study, precisely, when the flawed judgment was occurring. 

Another, we believe minor limitation, to interpreting our results is the narrowness of the integrity scale. As pointed out, the studies that donated the data only asked participants a portion of the questions in the integrity sub-scale. This was due to the original questions not having relevance for the HRI setting of the experiments. Obviously, changing domains already undermines the scale and warrants revalidation (arguably, new inventories need to be developed for human-agent interactions). Decreasing the number of questions in the inventory further exasperates this problem.

\section{Conclusions and Future Work}
We document instances of people providing ratings of agents that, rather than only considering information about the agent, reflect their experiences from a task in which they worked with the agent. For example, participants that did better in the task, all else being equal, tended to rate the benevolence of the agent higher. An implicit expectation of asking users to rate agents is that they will use only relevant information. These results point to people not following this normative expectation and using unrelated information in their evaluations. The traits in question, ability, benevolence, and integrity, are widely accepted as important components to trustworthiness. Thus, our findings suggest that the perceived trustworthiness of an agent, which is argued to impact compliance \cite{wang2016trust,gurney2022trust}, may be influenced by irrelevant information. 

The data that we analyze were not collected for this purpose; rather, they were collected to study how different degrees of explainability impact compliance. The measures were included out of a matter of practice; that is, their presence was not hypothesis-driven. Thus, the first step in future work is to undertake hypothesis-driven research in which experimental manipulation is used to study the correlation between attribute ratings and interaction outcomes. This includes using empirically validated measures for a broader range of traits in different empirical settings as well as formulating clearly stated hypotheses. One simple prediction is that positive (negative) interactions will lead to higher (lower) ratings of positive traits and lower (higher) ratings of negative traits. A related research question is: does the relevance of the trait to the interaction co-vary with the ratings? The prediction is that a highly important trait will receive more extreme ratings based on outcomes than a less relevant one.

Once these initial research questions are answered, the critical next step is exploring the cognitive processes driving the phenomenon. As noted, there are multiple plausible explanations, such as user affect, explanation-seeking behavior, and switching between causal and counterfactual reasoning based on outcomes---in all likelihood, it is some combination of these and other reasoning processes. Taking a controlled, experimental approach will allow us to understand how much variance each explanation may account for. We believe that the resulting insights will contribute to the development of more robust user models for deployment in agents.  

To illustrate, consider the hypothesis that a person uses their information about their own affective state when rating the agent's traits. This hypothesis factors down into two sub-predictions: first, there is the prediction that participants experience an affective response to the mission outcome, specifically, that they have a negative (positive) emotional response to doing poorly (well). Second, the contingent prediction, is that participants misuse this private information about their own affective state when rating the agent. Rather than allowing for performance and affect to vary naturally, we could fix performance such that all participants achieve similar results and induce different affective states, either related to the mission or not, and then collect assessments of the agent. This simple design would allow us to validate the misuse of affective information prediction as well as test the role of setting relevancy. Knowing how participants may (mis)use information about their own affective state may allow developers to anticipate and avoid the error. 

Another example of an aspect of the phenomenon that the data do not support exploring is whether priming participants about the eventual rating will alter how they view the interaction. To illustrate, telling participants before an interaction that they will eventually rate an agent's benevolence may change what information they ultimately use---it might even debias their responses such that they do not use irrelevant information, like their own affective state. Such debiasing experiments, once the variance in ratings is explained, are important steps towards curating predictable, appropriate interactions with intelligent agents. 

Perceived trustworthiness of AI agents is often cited as a determinant of successful human-agent interactions. Ideally, such perceptions would be based entirely on the features and behavior of the agent---not extemporaneous information. We believe that the effect documented herein, that a human's perception of an agent can change based on their own private information such as a feeling or emotion, points to an important aspect of human-agent interaction that should be more thoroughly studied. 
%% The acknowledgments section is defined using the "acks" environment
%% (and NOT an unnumbered section). This ensures the proper
%% identification of the section in the article metadata, and the
%% consistent spelling of the heading.
%\begin{acks}
%To whomever brings in the office snacks.
%\end{acks}

%%
%% The next two lines define the bibliography style to be used, and
%% the bibliography file.
\bibliographystyle{ACM-Reference-Format}
\bibliography{bib}

\end{document}